\documentstyle[twoside,fleqn,espcrc2,epsf]{article}

\newcommand{\AmS}{{\protect\the\textfont2
  A\kern-.1667em\lower.5ex\hbox{M}\kern-.125emS}}

\title{Potentials between heavy-light mesons from lattice and
       inverse scattering theory
\thanks{Supported in part by the Research Group in Physics of the
Hungarian Academy of Sciences, Debrecen,
by FWF P10468-PHY and by NSF PHY-9409195}
}

\author{H.R. Fiebig\address{Physics Department, 
                            FIU - University Park, 
                            Miami, Florida 33199, USA},
        H. Markum\address{Institut f\"ur Kernphysik,
                            Technische Universit\"at Wien,
                            A-1040 Vienna, Austria},
        A. Mih\'aly\address{Department of 
                            Theoretical Physics, 
                            Lajos Kossuth University,
                            H-4010 Debrecen, Hungary}
        and 
        K. Rabitsch$^{\mbox{\scriptsize b}}$
                                 }
              
\begin{document}

\begin{abstract}
We extend our investigation of heavy-light meson-meson
interactions to a system consisting of a heavy-light meson and the
corresponding antiparticle. An effective potential is obtained 
from meson-antimeson Green-functions computed in a quenched simulation
with staggered fermions. Comparisons with 
a simulation using an $O(a^2)$ tree-level and tadpole 
improved gauge action and a full QCD simulation show that
lattice discretization errors and dynamical quarks have no drastic influence.
Calculations from inverse scattering theory propose a similar 
shape for $K\bar{K}$ potentials.
\end{abstract}

\maketitle


Motivated by the success of QCD
in high energy scattering, today one aim is to 
calculate hadron-hadron interactions directly from the field equations of QCD.
In the low energy regime of QCD nonperturbative tools have to be used. 
This leads us to study the forces in systems of two hadrons on the lattice.

Previous lattice calculations have revealed a short range attractive
potential between two identical heavy-light color 
singlets \cite{Mih97}.  In this paper we present a lattice-QCD study of a
heavy-light meson-antimeson system $M\bar{M}$ 
and compare the resulting
residual interaction with $K\bar{K}$
potentials obtained from inverse scattering theory.


We define the one-meson field as a product of staggered Grassmann fields 
$\chi$ and $\bar{\chi}$ with a heavy and a light external flavor
$h$ and $l$, respectively,
\begin{equation}
\phi_{\vec{x}}(t) = \bar{\chi}_h(\vec{x}t) \chi_l(\vec{x}t) \,.
\label{eq1}\end{equation}
The meson-antimeson fields with relative distance $\vec{r}=\vec{y}-\vec{x}$ 
are then constructed by
\begin{equation}
\Phi_{\vec{r}}(t) = V^{-1} \sum_{\vec{x}} 
\sum_{\vec{y}} \delta_{\vec{r},\vec{y}-\vec{x}} \phi_{\vec{x}}(t)
 \phi^\dagger_{\vec{y}}(t) \,. 
\label{eq3}\end{equation}

The dynamics of the $M\bar{M}$ system is
contained in the time correlation matrix
\begin{equation}
C_{\vec{r}\vec{r}\,'}(t,t_0) = 
\langle \Phi^{\dagger}_{\vec{r}}(t) \Phi_{\vec{r}\,'}(t_0) \rangle 
 - \langle \Phi^{\dagger}_{\vec{r}}(t) \rangle 
   \langle \Phi_{\vec{r}\,'}(t_0) \rangle , 
\label{eq4}\end{equation}
where $\langle \; \rangle$ denotes the gauge field configuration average. 
On the hadronic level $C$ is a two-point correlator of a 
composite local operator describing a molecule-like structure.
Working out the contractions between the Grassmann fields we obtain
\begin{eqnarray}
 C_{\vec r}(t,t_0) &=&  V^{-2} \langle \sum_{\vec{x}}
\mbox{\rm Tr} (G^{(h)\dagger}_{\vec xt,\vec xt_0}G_{\vec xt,\vec xt_0}) 
\nonumber \\[-1mm] & \times &
\mbox{\rm Tr} (G^{(h)}_{\vec x+\vec rt,\vec x+\vec rt_0}
G^\dagger_{\vec x+\vec rt,\vec x+\vec rt_0}) \rangle 
+ \dots \nonumber \\[2mm]
&=& C^{(A)} + \, \dots \,.
\label{eq5}
\end{eqnarray}
Since the heavy valence quarks are fixed in space
the relative distance between the mesons 
is the same at the initial and final time of the propagation, 
$\vec{r}\,'=\vec{r}$. The heavy-quark propagator is
\begin{equation}
G^{(h)}_{\vec xt,\vec xt_0} = \left(\frac{1}{2m_{h}a}\right)^k
[\Gamma_{\vec x 4}]^k
\prod_{j=1}^kU_{x=(\vec x,ja),\mu =4} \, ,
\label{eq6}
\end{equation}
where the phase factors $\Gamma_{\vec x 4}=(-1)^{(x_1+x_2+x_3)/a}$
in the Kogut-Susskind formulation
correspond to the Dirac matrices and $k=(t-t_0)/a$.
In our calculations we set $m_{h}a=0.5$. The propagator of the light quark
is obtained from inverting the staggered
fermion matrix with a random source estimator.
A standard conjugate gradient algorithm is used.

The effective ground-state energy of the $M\bar{M}$ system 
$W(\vec r)$ can be extracted from the large euclidean
time behavior of $C$ following quantum-mechanical
reasoning for composite particles,
\begin{eqnarray}
C_{\vec{r}}(t,t_0)
& \simeq & c(\vec{r}\,) e^{-W(\vec{r})(t-t_0)} \,.
\label{eq7}
\end{eqnarray}
The residual meson-antimeson potential is
\begin{equation}
V(r)=W(r)-2m \, ,
\end{equation}  
where  $2m$ is the mass of two free (anti)mesons.


The simulations were performed on a periodic $8^3\times16$ lattice.
Each potential is the 
result of a measurement on 100 independent gauge field configurations
separated by 200 sweeps.
The inversion of the fermion matrix was performed with 32 random sources.

Each contribution to the correlator in 
(\ref{eq5}) comprises the exchange of gluons.  
As a first step only the direct term $C^{(A)}$
was computed, the computation of the full correlator 
being the task of a subsequent work.
The effective energy $W(r)$ was then extracted from the correlator $C^{(A)}$
by a four parameter (Levenberg-Marquardt) fit with the function
\begin{eqnarray}
C^{(A)}_r & = & B(r)\cosh{[ W(r)(t-8a)]}  \nonumber\\
 & + & (-1)^{t/a}
\widetilde{B}(r)\cosh{[ \widetilde{W}(r)(t-8a)]} .
\end{eqnarray}
The second term alternating in sign is a peculiarity of the
staggered scheme. To be numerically consistent with $W(r)$ at large 
distances, the mass $2m$
of two noninteracting mesons was extracted from the square of
the {\em one}-meson two-point function.

\begin{figure}[tb]
\centerline{ \epsfxsize 6.5cm \epsfbox{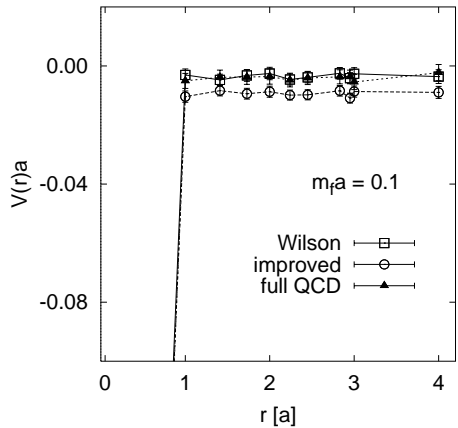}
}

\vspace{-0.8cm}

\caption{Heavy-light meson-antimeson
potentials $V(r) a$ for quark mass $m_fa = 0.1$,
from a quenched simulation using the Wilson  gauge action with
$\beta = 5.6$, from an improved
simulation with $\beta_{pl} = 7.0$
corresponding to the same lattice constant $a\approx 0.2\,$fm,
and from a full QCD simulation with number of flavors $n_f = 3$ and
$\beta = 5.2$.} 
\label{fig1}
\end{figure}

\begin{figure*}[t]
%
%
\centerline{\epsfxsize 5.5cm\epsfbox{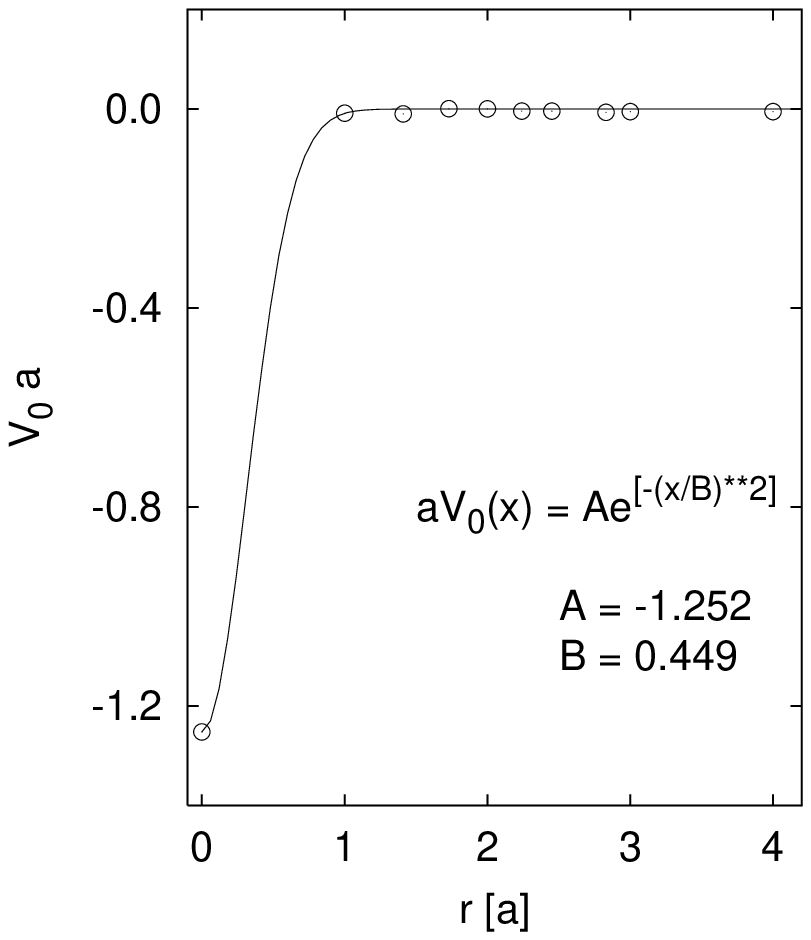}
\hspace{1.2cm} \epsfxsize 5.5cm \epsfbox{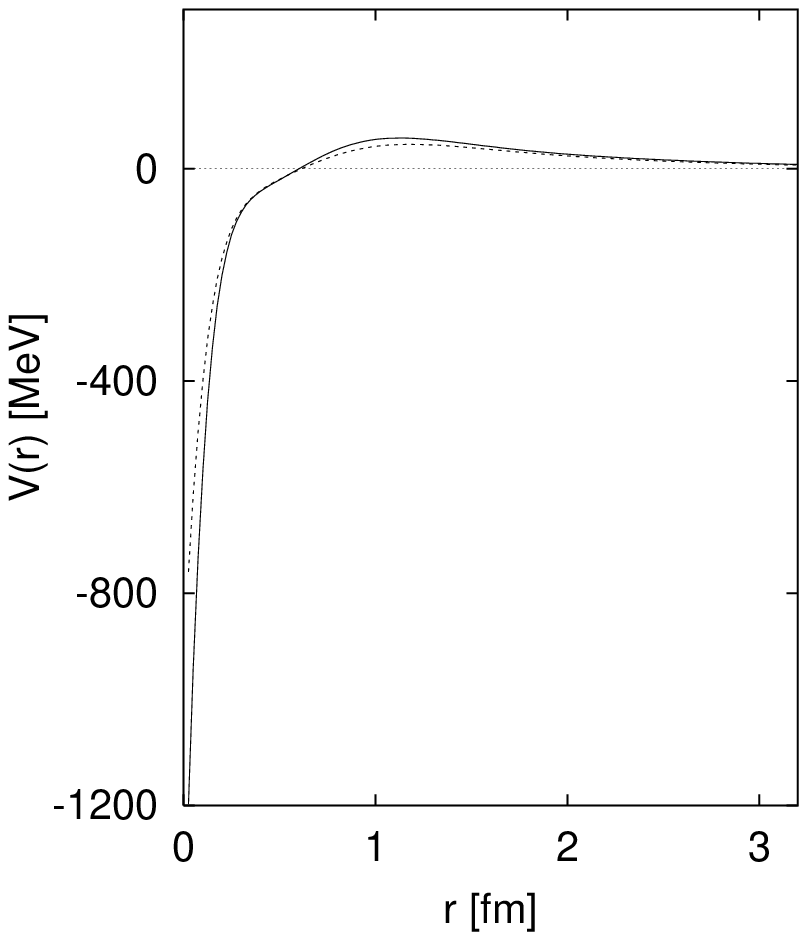}
}

\vspace{0.2cm}

\centerline{\epsfxsize 5.5cm\epsfbox{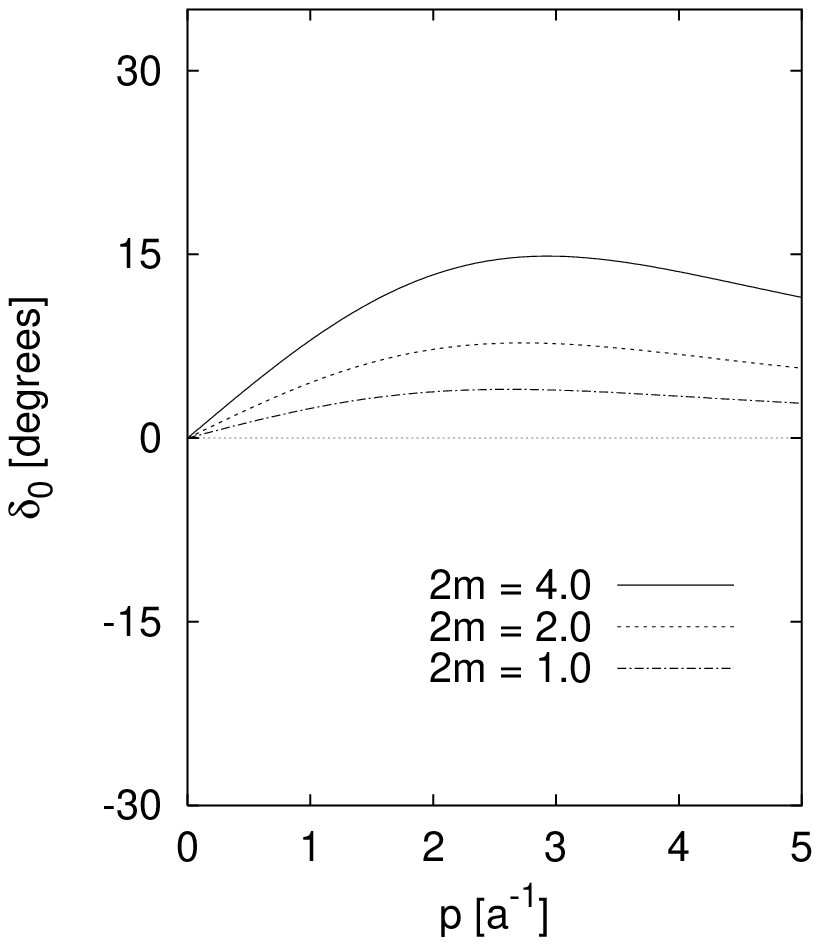}
\hspace{1.2cm} \epsfxsize 5.5cm \epsfbox{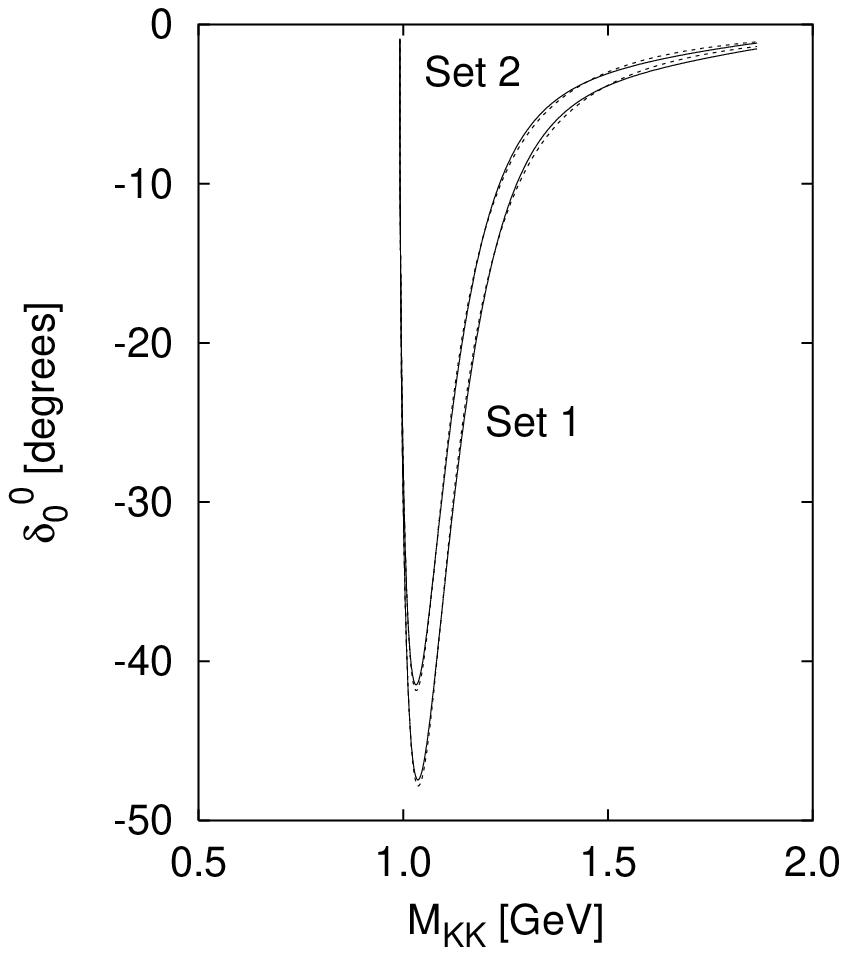}
}

\vspace{-.6cm}

\caption{Potentials and phase shifts for the $M\bar{M}$
system from the lattice (left plots)
and for the $K\bar{K}$ system
from inverse scattering (right plots).} 
\label{fig2}
\end{figure*}

In order to investigate the influence of discretization errors
and  sea quarks a simulation using an improved gauge action
and a full QCD simulation were
also performed. The former includes planar six-link plaquettes in
addition to the elementary plaquettes and is tadpole improved
\cite{alford}.
The resulting potentials are displayed in Fig.\ \ref{fig1}.
All simulations show consistent results. The potential turns out to
be short ranged and attractive. At distances $r/a \geq 1$ essentially no
interaction energy can be resolved.

Data from an independent simulation for a different light-quark mass $m_fa = 0.05$
were used to make a linear extrapolation to 
light quark mass zero. 
To estimate the strength of the interaction
the data were matched to a Gaussian potential.
The latter was then used to obtain scattering phase shifts
from a Schr\"{o}dinger equation with three arbitrary values
of the meson mass. The results are shown in 
the left plots of Fig.\ \ref{fig2}.

An effective range expansion with two sets of parameters
was applied to experimental data of
$K\bar{K}$ s-wave scattering \cite{kam94b}.
Using quantum inversion these analytic phase functions can be
transformed into local potentials \cite{San96}. The input phase shifts
and the resulting potentials are shown in 
the right plots of Fig.\ \ref{fig2}.
We notice a good qualitative agreement between lattice and inversion potentials
for short distances. 
The repulsive hump
could not be resolved by
our simulation. 
This feature, as it turns out, is decisive for
the phase shifts and explains the lack of qualitative
agreement between $\delta_0$.

The next step in our program is to 
compute the full correlator $C_r$. A  new simulation with an $O(a^2)$ 
tree-level and tadpole improved gauge action and Naik-type
improved staggered fermionic action is in progress. 
An independent lattice investigation with $O(a^2)$-improved lattice
action using Wilson fermions is presented in these proceedings.

\end{document}